\documentstyle[11pt]{article}
\def \bA {{\bar A}}
\def \bdel {{\bar \partial}}
\def \d {{\delta}}
\def \bd {\bar{\partial}}
\def \del {{\partial}}
\def \bD {{\bar D}}
\def \bz {{\bar z}}
\def \A {{\cal A}}
\def \E {{\cal E}}
\def \G {{\cal G}}
\def \H {{\cal H}}
\def \half {{\textstyle{1\over 2}}}
\def \la {{\langle}}
\def \O {{\cal O}}
\def \ra {{\rangle}}
\def \Tr {{\rm Tr}}
\def \vf {{\varphi}}

\def \vx {{\vec x}}
\def \by {{\bar y}}
\def \S {{\cal S}}

\newcommand{\be}{\begin{equation}}
\newcommand{\ee}{\end{equation}}
\newcommand{\no}{\nonumber}

\begin{document}
\begin{titlepage}
\null\vspace{-62pt}

\pagestyle{empty}
\begin{center}
\rightline{IASSNS-99/105}
\rightline{RU99-13-B}
\baselineskip=18pt
\vspace{1in}
{\Large \bf Gauge-invariant variables, WZW models and (2+1)-dimensional Yang-Mills theory}\\
\vspace{.5in}
V.P. NAIR \footnote{Permanent address: City College of the CUNY, New York, NY 10031}\\
\vskip .1in
{\it Institute for Advanced Study, Princeton, NJ 08540}\\
and\\
{\it Rockefeller University, New York, NY 10021\\
vpn@ajanta.sci.ccny.cuny.edu}\\
\end{center}
\vspace{1in}
\centerline{\bf Abstract}
Recent progress in understanding (2+1)-dimensional Yang-Mills ($YM_{2+1})$ theory via the use 
of gauge-invariant variables is reviewed. Among other things, we discuss
the vacuum wavefunction, an
analytic calculation of the string tension
and the propagator mass for gluons and its relation to 
the magnetic mass for $YM_{3+1}$ at nonzero temperature.
(Talk given at the Workshop on Physical Variables in Gauge Theories,
Dubna, Russia, September 1999.)
\end{titlepage}

\newpage
\pagestyle{plain}
\setcounter{page}{2}
\newpage
\baselineskip=14pt
\noindent{1. $\underline{Introduction}$}
\vskip .05in
In this talk I shall discuss a Hamiltonian approach to Yang-Mills theory in two spatial dimensions
($YM_{2+1}$), where nonperturbative calculations can be carried out to the extent that results
on mass gap and string tension can be compared with lattice simulations of the theory.
The work I shall report on was developed over the last few years in collaboration with 
D. Karabali and
Chanju Kim \cite{1,2,3}. (Eventhough I shall concentrate on the pure $YM_{2+1}$, some of the
results can be extended to the Yang-Mills-Chern-Simons theory \cite{4}).
Physical
variables or gauge-invariant variables play a key role in our analysis, in keeping 
with the theme of
this conference, and I hope this will be a nice example of how such variables can 
elucidate the
nonperturbative structure of gauge theories.

Before entering into the details of our work, let me say a few words about the relevance
of $YM_{2+1}$. Gauge theories (without matter) in (1+1) dimensions are rather trivial since
there are no propagating degrees of freedom, although there may be global degrees of
freedom on spaces of nontrivial topology. In (2+1) dimensions, gauge theories do have
propagating degrees of freedom and being next in the order of complexity, it is possible
that they provide a model simple enough to analyze mathematically and yet nontrivial
enough to teach us some lessons about (3+1)-dimensional
$YM$ theories. Another very important reason to study $YM_{2+1}$ is its relevance to
magnetic screening in $YM_{3+1}$ at high temperature. Gauge theories at finite temperature
have worse infrared problems than at zero temperature due to the divergent nature of the
Bose distribution for low energy modes. A dynamically generated Debye-type screening mass
will eliminate some of these, but we need a magnetic screening mass as well to have a
perturbative expansion which is well defined in the infrared.

A simple way to see that a magnetic mass can be dynamically generated
is as follows.
In the imaginary time
formalism, with Matsubara frequencies $\omega_n = 2\pi n T$, where $T$ is the
temperature, the gauge fields have a mode expansion as
$A_i(\vx ,x^0) = \sum_n A_{i,n}(\vx ) \exp ({i2\pi n T x^0})$. At high temperatures and for
modes of wavelength long compared to $1/T$, the modes with nonzero Matsubara frequencies are
unimportant and the theory reduces to the theory of the $\omega_n =0$ mode, viz., a
three (Euclidean) dimensional Yang-Mills theory (or a (2+1)-dimensional theory in
a Wick rotated version). Yang-Mills theories in three or (2+1) dimensions are expected to have
a mass gap and this is effectively the magnetic mass of the (3+1)-dimensional
theory at high temperature \cite{lingpy}. In order to incorporate this feature into
$YM_{3+1}$ at nonzero temperatures, one needs a decomposition of the $YM_{3+1}$ Feynman
integrals wherein the $YM_{2+1}$ pieces are isolated; in other words, one needs to
identify the ``slots" in the perturbative expansion of
$YM_{3+1}$ where the $YM_{2+1}$ results can be
inserted. There is a recent analysis along these lines by Reinbach and Schulz \cite{schulz}.

Let me now start by recalling a couple of facts about $YM_{2+1}$. The coupling constant
$e^2$ has the dimension of mass and it does not run as the four-dimensional
coupling does. The dimensionless expansion parameter of the theory is
$k/e^2$ or $e^2/k$, where $k$ is a typical momentum. Thus modes of low momenta
must be treated nonperturbatively, while modes of high momenta can be treated 
perturbatively. There is no simple dimensionless expansion parameter.
$YM_{2+1}$ is perturbatively super-renormalizable, so the ultraviolet  
singularities are well under control. 
\vskip .05in
\noindent{2. $\underline{The ~parametrization ~of ~the ~fields}$}
\vskip .05in
Coming now to the details of our analysis, let us consider 
a gauge theory with group $G=SU(N)$ in the $A_0 =0$ gauge. The gauge potential 
can be written as $A_i = -i t^a A_i ^a$, $i=1,2$, where $t^a$ are hermitian  
$N \times N$-matrices which form a basis of the Lie algebra of $SU(N)$ with 
$[t^a, t^b ] = i f^{abc} t^c,~~{\rm {Tr}} (t^at^b) = {1 \over 2} \delta ^{ab}$. 
The spatial coordinates $x_1 ,x_2$ will be combined into the complex combinations 
$z=x_1 -ix_2,~{\bar z} =x_1+ix_2$ with the corresponding components for the
potential 
$A\equiv A_{z} = {1 \over 2} (A_1 +i A_2), ~~  
{\bar A}\equiv A_{\bar{z}} = 
{1 \over 2} (A_1 -i A_2) = - (A_z)^{\dagger}$. 
The starting point of our analysis is a change of variables given by
\be 
A_z = -\partial_{z} M M^{-1},~~~~~~~~~~~~~ A_{\bar{z}} = M^{\dagger -1} \partial_ 
{\bar{z}} M^{\dagger}   
\label{1}
\ee 
Here $M,~M^\dagger$ are complex matrices in general, not unitary. If they are unitary,
the potenial is a pure gauge. 
The parametrization (\ref{1}) is possible
and is standard in many discussions of two-dimensional gauge fields. 
(There are also has similarities between (\ref{1}) and the construction of gauge-invariant
particle states as discussed by  McMullan, Lavelle and Horan at this 
workshop \cite{mcmullan}.)
A particular advantage of this parametrization is the way gauge transformations are
realized. A gauge transformation $A_i \rightarrow 
A_i^{(g)} = g^{-1} A_i g + g^{-1} \partial_i g, ~g(x)\in SU(N)$ is obtained 
by the transformation $M\rightarrow M^{(g)}=g M$. The gauge-invariant degrees of freedom
are parametrized by the hermitian matrix $H=M^\dagger M$.
Physical state wavefunctions are functions of $H$.

In making a change of variables in a Hamiltonian formalism, there are two things we
must do: 1) evaluate the volume measure (or Jacobian of the transformation) which
determines the inner product of the wavefunctions and 2) rewrite the Hamiltonian as an
operator involving the new variables. A consistency check would then be the
self-adjointness of the Hamiltonian with the given inner product. We begin with the
volume measure for the configuration space.
\vskip .05in
\noindent{3. $\underline{The ~functional ~measure ~and ~inner ~product}$}
\vskip .05in
The $YM$ Lagrangian in the $A_0 =0$ gauge is given by
\be
 L=  \int d^2x~ \left[ { e^2 \over 2 } {{\partial A_i^a} \over {\partial t}}  {{\partial
A_i^a}
\over {\partial t}}  -  {1 \over {2 e^2}} B^a B^a \right] 
\ee
where $B^a= \half \epsilon_{ij} (\partial_i A_j^a - \partial_j A_i^a +f^{abc}A_i^b
A_j^c)$. By comparison of the kinetic term with the standard point-particle Lagrangian
$L = \half g_{\mu \nu} {\dot q^{\mu} } {\dot q^{\nu}}$, 
we see that the metric for the fields $A, \bA$ is
$ds^2_{\cal A}~ =~ \int d^2x~ \delta A^a_i \delta A^a_i$, with the corresponding volume
$d\mu ({\cal A})= \prod_{x,a} dA^a (x) d\bA^a (x)$, which is the standard Euclidean volume.
From (\ref{1}) we see that
\begin{eqnarray}
\delta A&=& -D(\delta M M^{-1})= -\left( \partial (\delta M M^{-1} ) +[A, \delta M M^{-1}]\right)
\nonumber\\
\delta \bA &=& \bD (M^{\dag -1}\delta M^\dag ) \label{2}
\end{eqnarray}
which gives
\be
d\mu ({\cal A}) = (\det D \bD )~d\mu (M, M^\dag )
\label{3}
\ee
where $d\mu (M, M^\dag )$ is the volume for the complex matrices $M, M^\dag$, which is associated
with the metric $ds_M^2~=8 \int {\Tr}(\delta M
M^{-1}~M^{\dagger -1} \delta M^\dagger )$. This is given by the highest order differential form
$dV$ as $d\mu (M, M^\dag )= \prod_x dV(M,M^\dag )$ where
\begin{eqnarray}
dV(M,M^{\dagger}) \propto && \epsilon _{a_1...a_n}  (dM M^{-1})_{a_1}
\wedge ...  \wedge (dMM^{-1})_{a_n}  \nonumber\\
 &&~\times  \epsilon _{b_1...b_n} (M^{\dagger -1} d M^{\dagger})_{b_1} \wedge ...
\wedge  (M^{\dagger -1} d M^{\dagger})_{b_n}\label{4}
\end{eqnarray}
where $n={\rm dim} G={\rm dim} SU(N) = N^2 -1$. (There are some
constant numerical factors which are irrelevant for our discussion.)
The complex matrix $M$ can be written as $M= U \rho$, where $U$ is unitary and $\rho$ is hermitian.
This is the matrix analogue of the modulus and phase decomposition for a complex number.
Since gauge transformations act as $M\rightarrow M^{(g)}=g M$, we see that $U$ represents the gauge
degrees of freedom and $\rho$ represents the gauge-invariant degrees of freedom on $M$.
Substituting $M = U\rho$, (\ref{4}) becomes
\begin{eqnarray}
dV(M, M^{\dagger}) \propto  && \epsilon _{a_1...a_n} (d\rho \rho^{-1} +
\rho^{-1} d\rho)_{a_1} \wedge ... \wedge (d\rho \rho^{-1} + \rho^{-1} d\rho)_{a_n} \nonumber\\
&&~\times \epsilon _{b_1...b_n} (U^{ -1} d U)_{b_1} \wedge ... \wedge (U^{ -1} d
U)_{b_n} \nonumber\\
\propto && \epsilon _{a_1...a_n} (H^{-1}dH)_{a_1} \wedge ...  \wedge (H^{-1}dH)_{a_n}  
d\mu (U)\label{5}
\end{eqnarray}
Here $d\mu (U)$ is the standard group volume measure (the  Haar measure)
for $SU(N)$.
Upon taking the product over all points, $d\mu (U)$ gives the
volume of the entire gauge group (namely all $SU(N)$-valued functions) which we denote by
$vol(\G_*)$ and thus
\begin{eqnarray}
d\mu (M, M^{\dagger}) &&= \prod_{x} dV(M, M^{\dag}) 
~ vol (\G _*)  = d\mu (H)  ~vol (\G _*) \label{6}\\
d\mu (H)&& = \prod_{x,a}  \det r [d\vf ^a] \no \\
\det r\prod_a d\vf ^a&&= \epsilon _{a_1...a_n} (H^{-1}dH)_{a_1} ... (H^{-1}dH)_{a_n}  
\end{eqnarray}
We have parametrized $H$ in terms of the real
parameters $\vf ^a$ and $H^{-1}dH = d\vf ^a r_{ak} (\vf) t_k$. 
The volume element or the integration measure for the gauge-invariant configurations 
can now be written as
\begin{eqnarray}
{d\mu ({\A})\over vol({\G}_*)}
&&= {[dA_z dA_{\bar{z}}]\over vol({\G}_*)} \no \\
&&= (\det D_z D_{\bar{z}}) {d\mu  (M,
M^{\dagger})\over vol({\G}_*)} ~= (\det D \bD ) d\mu (H)\label{7}
\end{eqnarray}
where we have used (\ref{6}).
The problem is thus reduced to the calculation of the determinant of the
two-dimensional operator $D\bD$. This is well known \cite{poly}. 
The simplest way to evaluate this is to define $\Gamma = \log~\det D\bD$, which gives
\be
{\delta \Gamma \over \delta \bA^a}~= -i~\Tr\left[ \bD^{-1}(x,y) 
T^a\right]_{y\rightarrow x}\label{7b}
\ee
$(T^a)_{mn}=-if^a_{mn}$ are the generators  of the Lie algebra in the adjoint
representation. The coincident-point limit of $\bD^{-1}(x,y)$ is
singular and needs regularization. With a gauge-invariant regulator,one finds
\be
\Tr \left[ \bD^{-1}_{reg}(x,y) T^a \right]_{y\rightarrow x}~= {2c_A \over \pi}
\Tr \left[ (A -M^{\dag -1} \partial M^\dag )t^a\right]\label{7c}
\ee
where $c_A \delta^{ab} = f^{amn}f^{bmn}$; it is equal to $N$ for $SU(N)$.
Using this result in (\ref{7b}) and integrating we get 
\be
(\det D \bD) ~= \left[ {{\det ' \del \bdel } \over \int d^2 x} \right] ^{{\rm dim}
G} ~ \exp \left[ 2c_A ~\S (H) \right]
\label{8}
\ee
$\S (H)$ is the 
Wess-Zumino-Witten (WZW) action for the hermitian matrix field $H$ given by \cite{witt}
\be
{\S} (H) = {1 \over {2 \pi}} \int \Tr (\partial H \bar{\partial} H^{-1}) +{i
\over {12 \pi}} \int \epsilon ^{\mu \nu \alpha} \Tr ( H^{-1} \partial _{\mu} H H^{-1}
\partial _{\nu}H H^{-1} \partial _{\alpha}H) \label{9}
\ee

We can now write the inner product for states $|1\ra$ and
$|2\ra$, represented by the wavefunctions $\Psi_1$ and $\Psi_2$, as \cite{GKBN}
\be
\label{inprod}
\la 1 | 2\ra = \int d\mu (H) e^{2c_A ~\S (H)}~~\Psi_1^* \Psi_2 \label{10}
\ee
\vskip .05in
\noindent{4. $\underline{Transforming ~the ~Hamiltonian}$}
\vskip .05in
The next step is the change of variables in the Hamiltonian. However, 
there is some
further simplification we can do before taking up the Hamiltonian. 
We would expect the wavefunctions to be functionals of the matrix field
$H$, but actually we can take them to be functionals of the current of the WZW model
(\ref{9}) given by
$J= (c_A/\pi ) \partial_zH~H^{-1}$. First of all we notice that the Wilson loop operator
can be constructed from $J$ alone as
\be
W(C)= \Tr P ~e^{-\oint_C (Adz+\bA d{\bar z})}=
 \Tr P ~e^{(\pi /c_A)\oint_C J }\label{11}
\ee
Since the Wilson loop operator can provide a complete description of gauge-invariant 
observables, it is sufficient to take wavefunctions to be functions of $J$.
There is also a conformal theory argument to show that it is sufficient to
conside only functions of $J$ \cite{1,2}.

This means that we can transform the Hamiltonian ${\cal H}= T+V$
to express it in terms of
$J$ and functional derivatives with respect to $J$.
This is achieved by the chain rule of differentiation
\begin{eqnarray}
T\Psi &&= {e^2\over 2}\int E^a_iE^a_i ~\Psi\no\\
&&= -{e^2\over 2}\left[ \int_{x,u}{\delta  J^a ( u )\over \delta A^c_i(x)\delta A^c_i(x)}
{\delta \Psi
\over \delta J^a(u)} + \int_{x,u,v}{\delta J^a(u) \over \delta A^c_i(x)}{\delta J^b(v)\over \delta
A^c_i(x)} ~{\d \over \d J^a( u) }{\d \over \d J^b(v) }\Psi \right]\no\\
V&&={1\over 2e^2} \int B^aB^a  \label{12}
\end{eqnarray}
Regularization is important in calculating the coefficients of the two terms in $T$.
Carrying this out we find
\begin{eqnarray}
T&&= m \left[ \int_u J^a(u) {\d \over \d J^a(u)}~+~ \int \Omega_{ab} (u,v) 
{\d \over \d J^a(u) }{\d \over \d J^b(v) }\right] \label{13a}\\
V&&={ \pi \over {m c_A}} \int \bdel J_a (\vx)
\bdel J_a (\vx) \label{13b}
\end{eqnarray}
where $m= e^2c_A/2\pi$ and
\be
\Omega_{ab}(u,v) = {c_A\over \pi^2} {\d_{ab} \over (u-v)^2} ~-~ 
i {f_{abc} J^c (v)\over {\pi (u-v)}}
\ee
The first term in $T$ shows that every power of $J$ in the wavefunction gives
a value $m$ to the energy, suggesting the existence of a mass gap.
The calculation of this term involves exactly the same quantity as in (\ref{7b})
and with the same regulator leads to (\ref{13a}), i.e.,
\begin{eqnarray}
-{e^2\over 2}\int d^2y {\delta^2 J_a(x)\over 
\delta \bA^b(y) \delta A^b(y)} &&={e^2c_A\over 2\pi} M^{\dag} _{am} \Tr \left[ T^m
\bD ^{-1}(y,x) 
\right]_{y \rightarrow x}\no\\
&&= m ~J_a(x)\label{13c}
\end{eqnarray}

Finally, (\ref{13a},\ref{13b}) (with regularizations taken account of) give a self-adjoint Hamiltonian
which, as I mentioned before, is a nice consistency check.
\vskip .05in
\noindent{5. $\underline{An ~intuitive ~argument}$}
\vskip .05in
The next step is to solve the Schr\"odinger equation, at least for the vacuum state 
and the low lying
excited states. However, before taking this up, I shall
give a short intuitive argument for the existence of a mass gap.

First of all notice that the total volume of the configuration space as given by
(\ref{7}) is finite,
modulo regularization of the Laplacian
$\partial \bar{\partial}$, and can be written as \cite{GKBN,1}
\be
\int {[dA_z dA_{\bar{z}}]\over vol({\G}_*)}=\left[ {{({\rm det}' \partial \bar{\partial})} 
\over {\int
d^2x}} \right] ^ {-{\rm dim}G} \label{vol}
\ee
In contrast to this, the corresponding result
for an Abelian theory (which has $c_A=0$) is infinite. 
(If we make a mode decomposition of $H$ or $\vf^a$ over the eigenmodes
of the Laplacian $\partial \bdel$, the integration over the amplitude of 
each mode is finite for the nonabelian case because of
the exponential; the divergence
arises from the infinity of modes and can be regularized by truncation
to a finite number of modes. For the Abelian case, the integration for each
mode is divergent.) This result is encouraging as regards the question of the mass gap.
One can go further and make a slightly better argument.
The crucial
ingredient is the measure of integration in the inner product
(\ref{10}). 
Writing $\Delta E, ~\Delta B$ for the root mean square fluctuations of the  
electric field $E$ 
and the magnetic field 
$B$, we have, from the canonical commutation rules  
$[E_i^a, A_j^b]= -i\delta_{ij}\delta^{ab}$, $\Delta E~\Delta B\sim k$,  
where $k$ is the 
momentum variable. This gives an estimate for the energy 
\be 
{\E}={1\over 2} \left( {e^2 k^2\over\Delta B^2 } +{\Delta B^2 \over e^2} 
 \right) 
\ee 
For low lying states, we minimize ${\E}$ with respect to $\Delta B^2$,  
$\Delta B^2_{min}\sim 
e^2 k$, giving ${\E}\sim k$. This corresponds to the standard photon. 
For the nonabelian theory, this is inadequate since $\la \H \ra$ 
involves 
the factor $e^{2c_A \S (H) }$. In fact, 
\be
\la \H \ra ~\sim \int d\mu (H) e^{2c_A\S (H) }~ \half (e^2E^2 +B^2/e^2 ) 
\ee 
In terms of $B$, the WZW action goes like $ \S (H) \approx 
[ -(c_A /\pi ) \half \int B (1/k^2 )B +...]$; we thus see that $B$ follows a Gaussian  
distribution of width $\Delta B^2 \approx \pi k^2 /c_A$, for  
small values of $k$. This Gaussian dominates near small $k$ 
giving $\Delta B^2 \sim k^2 (\pi /c_A )$.  
In other words, eventhough ${\E}$ is minimized around $\Delta B^2 \sim k$, 
probability is  
concentrated around $\Delta B^2 \sim k^2 (\pi /c_A)$. For the expectation 
value of the energy, 
we then find 
${\E}\sim e^2c_A/2\pi  +{\O}(k^2)$. Thus the kinetic term in  
combination with  
the measure factor $e^{2c_A\S (H)}$ could lead to a mass gap of order $e^2c_A$.  
The argument is not rigorous, but captures the essential physics as we shall see
in a moment.
\vskip .05in
\noindent{6. $\underline{The ~vacuum ~wavefunction}$}
\vskip .05in
Let us now consider the eigenstates of the theory.
The vacuum wavefunction is presumably the simplest to calculate.
Ignoring the potential term $V$ for the moment, since $T$ involves derivatives, 
we see immediately that
the ground state wavefunction for $T$ is $\Phi_0 =1$. This may seem like a 
trivial statement, but the key
point is that it is normalizable with the inner product (\ref{10}); 
in fact, the normalization integral is
(\ref{vol}). Starting with this, we can solve the Schro\"odinger equation
taking $\Psi_0$ to be of the form $\exp (P)$, where $P$ is a perturbative series in the potential
term $V$ (equivalent to a $1/m$-expansion). We then get
\begin{eqnarray}
P = && - {\pi \over { m^2 c_A}} \Tr \int  : \bdel J \bdel J : \no\\
&& - \left({\pi \over { m^2 c_A}}\right)^2 \Tr \int   \bigl[: \bdel J ( {\cal D} \bd ) \bdel J 
    +  {1 \over 3} \bdel J  [J, \bdel ^2 J] : \bigr] \no\\  
&& - 2 \left({\pi \over {m^2 c_A}}\right)^3 \Tr \int \bigl[ : \bdel J  ( {\cal D} \bd )^2 \bd J 
+{2 \over 9} [ {\cal D} \bd J,~ \bd J] \bd ^2 J + {8 \over 9} [{\cal D} \bd ^2
J,~ J] \bd ^2 J \no\\
&&~~~~~~~~~~~~~~~~- {1 \over 6} [J, ~ \bd J] [\bd J,~ \bd ^2 J] - {2 \over 9} [J,
\bd J][J, \bd ^3 J]: \bigr] 
+ {\cal O} ( {1 \over m^8})\label{P}
\end{eqnarray}
where ${\cal D}h= ({c_A / \pi}) \del h -[J,h]$.
The series is naturally grouped as terms with 2 $ J$'s, terms with 3 $J$'s, etc.
These terms can be summed up; for the $2J$-terms we find
\be
\Psi_0= \exp\left[ -{1 \over {2 e^2}} \int_{x,y} B_a(x) \left[{ 1 \over  {\bigl( m + 
\sqrt{m^2 - \nabla ^2 } \bigr)} }\right] _{x,y} B_a(y) ~+ {\cal O}(3J)\right]
\label{wavfn}
\ee
The first term in (\ref{wavfn}) has the correct (perturbative) high momentum
limit, viz.,
\be
\Psi_0\approx  \exp\left[ -{1 \over {2 e^2}} \int_{x,y} B_a(x) \left[{ 1 \over
\sqrt{ - \nabla ^2 } }\right] _{x,y} B_a(y) ~+ {\cal O}(3J)\right]
\label{wavfn2}
\ee
Thus although we started with the high $m$ (or low momentum)
limit, the result (\ref{wavfn}) does match onto the perturbative limit.
The higher terms are also small for the low momentum limit.

We can now use this result to calculate the expectation value of the
Wilson loop operator; for the fundamental representation,
it is given by
\begin{eqnarray}
\la W_F (C) \ra &&= {\rm constant}~~\exp \left[ - \sigma {\cal
A}_C \right]\no\\
{\sqrt{\sigma }}&&= e^2 \sqrt{{N^2-1\over 8\pi}}
\label{tension}
\end{eqnarray}
where ${\cal A}_C$ is the area of the loop $C$. $\sigma$ is the string tension.
This is a prediction of our analysis starting from first principles with no adjustable
parameters. Notice that the dependence on $e^2$ and $N$ is
in agreement 
with large-$N$ expectations, with $\sigma$ depending only on the combination
$e^2N$ as $N\rightarrow \infty$. (The first correction to the large-$N$ limit
is negative, viz., $-(e^2N)/2N^2\sqrt{8\pi}$ which may be interesting in the context
of large-$N$ analyses.)
Formula (\ref{tension}) gives the values $\sqrt{\sigma}/e^2
=0.345, 0.564, 0.772, 0.977$ for $N=2,3,4,5$. 
There are estimates for $\sigma$ based on Monte Carlo simulations of lattice gauge
theory. The most recent results for the gauge  groups $SU(2),~SU(3),
~SU(4)$ and $SU(5)$ are 
$\sqrt{\sigma}/e^2 =$ 0.335, 0.553, 0.758, 0.966 \cite{teper}. We see that
our result agrees with the lattice result to within $\sim 3\%$.

One might wonder at this stage why the result is so good when we have
not included the $3J$- and higher terms in the the wavefunction.
This is basically because the
string tension is determined by large area loops and for these, it is the
long  distance part of the wavefunction which contributes significantly.
In this limit, the $3J$- and higher terms in (\ref{wavfn}) are small compared to the
quadratic term. 

We have summed up the $3J$-terms as well. Generally, one finds that $P$,
when expressed in terms of the magnetic field, is nonlocal
even in a $(1/m)$-expansion, contrary
to what one might expect for a theory with a mass. This is essentially due to 
gauge invariance combined with our choice of $A_0=0$; it has recently been shown that
a similar result holds for the Schwinger model \cite{mansfield}.
\vskip .05in
\noindent{7. $\underline{Magnetic ~mass}$}
\vskip .05in
I shall now briefly return to the magnetic mass. From the expression (\ref{13a})
we see immediately that for a wavefunction which is just $J^a$, we have the exact
result $T ~J^a = m ~J^a$. 
When the potential term is added, $J^a$ is no longer an exact eigenstate; we find
\be
(T+V) ~J^a = \sqrt{m^2 -\nabla^2}~ J^a ~+~\cdots
\ee
showing how the mass value is corrected to the relativistic dispersion relation.

Now $J^a$ may be considered as the gauge-invariant definition
of the gluon. This result thus suggests a dynamical propagator
mass $m= e^2c_A/2\pi$ for the gluon.
A different way to see this result is as follows.
We can expand the matrix field $J$ in powers of $\vf_a$ which parametrizes $H$, 
so that
$J\simeq (c_A/\pi ) \partial \vf_a t_a$. This is like a perturbation expansion, but
a resummed or improved version of it, where we expand the WZW action in
$\exp(2c_A\S (H))$ but not expand the exponential itself. The Hamiltonian can then be
simplified as
\be
\label{hamil2}
\H \simeq \half \int_x [- {\d ^2 \over {\d \phi _a ^2 (x)}} + \phi_a (x)  \bigl( m^2 -
\nabla ^2 \bigr)  \phi_a (x)] + ...
\ee
where $\phi_a (k) \!=\! \sqrt {{c_A k \bar{k} }/ (2 \pi m)} \vf _a (k)$, in momentum space.
In arriving at this expression we have expanded the currents and also absorbed the
WZW-action part of the measure into the definition of the wavefunctions, i.e.,
the operator (\ref{hamil2}) acts on 
${\tilde \Psi} =e^{c_AS(H)}\Psi$. 
The above equation shows that the propagating particles in the
perturbative regime, where the power series expansion of the current is
appropriate, have a mass $m=e^2c_A/2\pi$. 
This value can therefore be identified as the magnetic
mass of the gluons as given by this  nonperturbative analysis.

For $SU(2)$ our result is $m\approx 0.32 e^2$.  There have been two lattice calculations
of the propagator mass of gluons for $SU(2)$; the values are $m\approx 0.35e^2$ and
$0.46e^2$ \cite{karsch1,karsch2}. 
There is reasonable evidence for a `constituent gluon' picture for
glueballs in $YM_3$ from lattice analysis, so another suggestion has been to extract
a constituent mass for a gluon and interpret it as the magnetic mass \cite{owe}. This gives a
value $m\approx 0.31 e^2-0.40 e^2$. Considering the difficulties of a lattice estimate and
the variance within these calculations, we cannot draw any definite conclusion in
comparing with our analysis.
\vskip .05in
\noindent{8. $\underline{Excited ~states}$}
\vskip .05in
Eventhough $J$ is useful as  a description of the gluon, it is not a physical state.
This is because of an ambiguity in our parametrization (\ref{1}). Notice that
the matrices $M$ and $M{\bar V}(\bz )$ both give the same $A, \bA$, where
${\bar V}(\bz )$ only depends on $\bz$ and not $z$.  Since we have the same potentials,
physical results must be insensitive to this redundancy in the choice of $M$;
in other words, physical wavefunctions must be invariant under $M\rightarrow
M {\bar V}(\bz )$. $J$ is not invariant; we need at least two $J$'s to form
an invariant combination. An example is
\be
\Psi _2 = \int _{x,y} f(x,y) \bigl[ \bdel J_a (x) \bigl( H(x,\by) H^{-1} (y,
\by) \bigr) _{ab} \bdel J_b (y)  \bigr]
\ee
This is not an eigenstate of the Hamiltonian.
Since there are two $J$'s we should expect at least a mass
of $2m$, but beyond that it is difficult to say anything very conclusive, see 
however \cite{2}.
\vskip .2in
This work was supported in part by a grant from the National Science Foundation.


\end{document}